# Negative local resistance due to viscous electron backflow in graphene

D. A. Bandurin[1], I. Torre[2,3], R. Krishna Kumar[1,4], M. Ben Shalom[1,5], A. Tomadin[6], A. Principi[7], G. H. Auton[5], E. Khestanova[1,5], K. S. Novoselov[5], I. V. Grigorieva[1], L. A. Ponomarenko[1,4], A. K. Geim[1], M. Polini[3]

[1]School of Physics & Astronomy, University of Manchester, Oxford Road, Manchester M13 9PL, United Kingdom
[2]National Enterprise for nanoScience and nanoTechnology, Scuola Normale Superiore, I-56126 Pisa, Italy
[3]Istituto Italiano di Tecnologia, Graphene labs, Via Morego 30 I-16163 Genova (Italy)
[4]Physics Department, Lancaster University, Lancaster LA14YB, United Kingdom
[5]National Graphene Institute, University of Manchester, Manchester M13 9PL, United Kingdom
[6]National Enterprise for nanoScience and nanoTechnology, Istituto Nanoscienze-Consiglio Nazionale delle Ricerche and Scuola Normale Superiore, I-56126 Pisa, Italy
[7]Radboud University, Institute for Molecules and Materials, NL-6525 AJ Nijmegen, The Netherlands

**Graphene hosts a unique electron system in which electron-phonon scattering is extremely weak but electron-electron collisions are sufficiently frequent to provide local equilibrium above liquid nitrogen temperature. Under these conditions, electrons can behave as a viscous liquid and exhibit hydrodynamic phenomena similar to classical liquids. Here we report strong evidence for this long-sought transport regime. In particular, doped graphene exhibits an anomalous (negative) voltage drop near current injection contacts, which is attributed to the formation of submicrometer-size whirlpools in the electron flow. The viscosity of graphene's electron liquid is found to be ≈0.1 m$^2$s$^{-1}$, an order of magnitude larger than that of honey, in agreement with many-body theory. Our work shows a possibility to study electron hydrodynamics using high quality graphene.**

Collective behavior of many-particle systems that undergo frequent inter-particle collisions has been studied for more than two centuries and is routinely described by the theory of hydrodynamics (*1,2*). The theory relies only on the conservation of mass, momentum and energy and is highly successful in explaining the response of classical gases and liquids to external perturbations varying slowly in space and time. More recently, it has been shown that hydrodynamics can also be applied to strongly interacting quantum systems including ultra-hot nuclear matter and ultra-cold atomic Fermi gases in the unitarity limit (*3-6*). In principle, the hydrodynamic approach can also be employed to describe many-electron phenomena in condensed matter physics (*7-13*). The theory becomes applicable if electron-electron scattering provides the shortest spatial scale in the problem such that $\ell_{\rm ee} \ll W, \ell$ where $\ell_{\rm ee}$ is the electron-electron scattering length, $W$ the characteristic sample size, $\ell \equiv v_{\rm F}\tau$ the mean free path, $v_{\rm F}$ the Fermi velocity, and $\tau$ the mean free time with respect to momentum-non-conserving collisions such as those involving impurities, phonons, etc. The above inequalities are difficult to meet experimentally. Indeed, at low temperatures ($T$) $\ell_{ee}$ varies approximately as $\propto T^{-2}$ reaching a micrometer scale at liquid-helium $T$ (*14*), which necessitates the use of ultra-clean systems to satisfy $\ell_{ee} \ll \ell$. At higher $T$, electron-phonon scattering rapidly reduces $\ell$. However, for two-dimensional (2D) systems with dominating acoustic phonon scattering, $\ell$ decays only as $\propto T^{-1}$, slower than $\ell_{ee}$, which should in principle allow the hydrodynamic description over a certain temperature range, until other phonon-mediated processes become important. So far, there has been little evidence for hydrodynamic electron transport. An exception is an early work on 2D electron gases in ballistic devices ($\ell \sim W$) made from GaAlAs heterostructures (*15*). They exhibited nonmonotonic changes in differential resistance as a function of a large applied current $I$ that was used to increase the electron temperature (making $\ell_{ee}$ short) while the lattice temperature remained low (allowing long $\ell$). The nonmonotonic behavior was attributed to the Gurzhi effect, a transition between Knudsen ($\ell_{ee} \gg \ell$) and viscous electron flows (*7,15*). Another possible hint for electron hydrodynamics comes from one of the explanations (*16,17*) for the Coulomb drag measured between two graphene sheets at the charge neutrality point (CNP).

Here we address electron hydrodynamics by using a special measurement geometry (Fig. 1) that amplifies effects of the shear viscosity $\nu$ and, at the same time, minimizes a contribution from ballistic effects that can occur not only in the Knudsen regime but also for viscous flow in graphene. As shown in Figs. 1A-B, a viscous flow can lead to vortices appearing in the spatial distribution of the steady-state current. Such 'electron whirlpools' have a spatial scale $D_\nu = \sqrt{\nu\tau}$, which depends on electron-electron scattering through $\nu$ and on the electron system's quality through $\tau$ (*18*). To detect the whirlpools, electrical probes should be placed at a distance comparable to $D_\nu$. By using single- and bi- layer graphene (SLG and BLG, respectively) encapsulated between boron nitride crystals (*19-21*), we could reach $D_\nu$ of 0.3-0.4 µm due to high viscosity of graphene's Fermi liquid and its high carrier mobility $\mu$ even at high $T$. To the best of our knowledge, such large $D_\nu$ are unique to graphene but still necessitate submicron resolution to probe the electron backflow. To this end, we fabricated multiterminal Hall bars with narrow ($\approx 0.3$ µm) and closely spaced ($\approx 1$ µm) voltage probes (Fig. 1C and *fig. S1*). For details of device fabrication, we refer to Supporting Material (*18*).

All our devices were first characterized in the standard geometry by applying $I$ along the main channel and using side probes for voltage measurements. A typical behavior of longitudinal conductivity $\sigma_{xx}$ at a few characteristic $T$ of interest is shown in Fig. 1D. At liquid-helium $T$, the devices exhibited $\mu \sim$ 10-50 m$^2$V$^{-1}$s$^{-1}$ over a wide range of carrier concentrations $n \sim 10^{12}$ cm$^{-2}$, and $\mu$ remained above 5 m$^2$V$^{-1}$s$^{-1}$ up to room $T$ (*fig. S2*). Such $\mu$ allow ballistic transport with $\ell > 1$ µm at $T < 300$ K. On the other hand, at $T \geq 150$ K $\ell_{ee}$ decreases down to 0.1-0.3 µm over the same range of $n$ (*22, 23* and *figs. S3-S4*). This allows the essential condition for electron hydrodynamics ($\ell_{ee} \ll W, \ell$) to be satisfied within this temperature range. If one uses the conventional longitudinal geometry of electrical measurements, it turns out that viscosity has little effect on $\sigma_{xx}$ (*figs. S5-S7*) essentially because the flow in this geometry is uniform whereas the total momentum of the moving Fermi liquid is conserved in electron-electron collisions (*18*). The only evidence for hydrodynamics we could find in the longitudinal geometry was the Gurzhi effect that appeared as a function of the electron temperature controlled by applying large $I$, similar to the observations of ref. *15* (*fig. S8*).

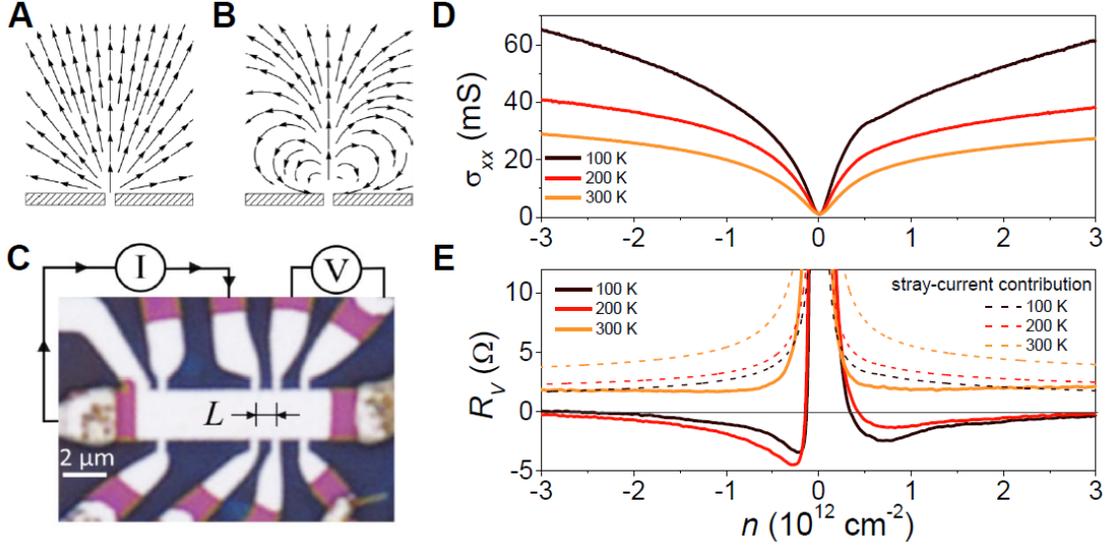

**Fig. 1. Viscous backflow in doped graphene.** (**A,B**) Calculated steady-state distribution of current injected through a narrow slit for a classical conducting medium with zero $\nu$ (A) and a viscous Fermi liquid (B). (**C**) Optical micrograph of one of our SLG devices. The schematic explains the measurement geometry for vicinity resistance. (**D,E**) Longitudinal conductivity $\sigma_{xx}$ and $R_V$ as a function of $n$ induced by applying gate voltage. $I = 0.3$ µA; $L = 1$ µm. The dashed curves in (E) show the contribution expected from classical stray currents in this geometry (*18*).

To reveal hydrodynamics effects, we employed the geometry shown in Fig. 1C. In this case, $I$ is injected through a narrow constriction into the graphene bulk, and the voltage drop $V_V$ is measured at the nearby side contacts located at the distance $L \sim 1$ µm away from the injection point. This can be considered as nonlocal measurements, although stray currents are not exponentially small (dashed curves in Fig. 1E). To distinguish from the proper nonlocal geometry (*24*), we refer to the linear-response signal measured in our geometry as "vicinity resistance", $R_V = V_V/I$. The idea is that, in the case of a viscous flow, whirlpools emerge as shown in Fig. 1B, and their appearance can then be detected as sign reversal of $V_V$, which is positive for the conventional current flow (Fig. 1A) and negative for viscous backflow (Fig. 1B). Fig. 1E shows examples of $R_V$ for the same SLG device as in Fig. 1D, and other SLG and BLG devices exhibited similar behavior (*18*). One can see that, away from the CNP, $R_V$ is indeed negative over a wide range of intermediate $T$, despite a significant offset expected due to stray currents. Figure 2 details our observations further by showing maps $R_V(n,T)$ for SLG and BLG. The two Fermi liquids exhibited somewhat different behavior reflecting their different electronic spectra but $R_V$ was negative over a large range of $n$ and $T$ for both of them. Two more $R_V$ maps are provided in *fig. S9*. In total, seven multiterminal devices with $W$ from 1.5 to 4 µm were investigated showing the vicinity behavior that was highly reproducible for both different contacts on a same device and different devices, independently of their $W$, although we note that the backflow was more pronounced for devices with highest $\mu$ and lowest charge inhomogeneity.

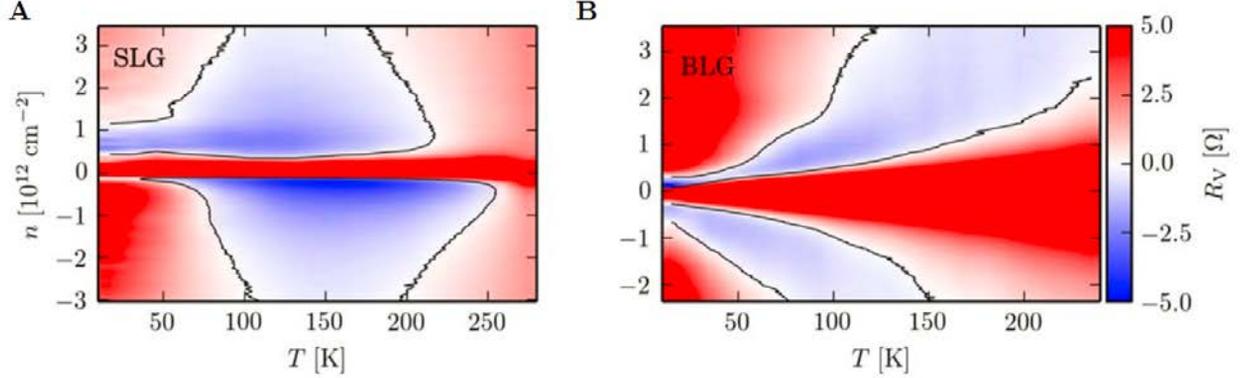

**Fig. 2. Vicinity resistance maps. (A, B)** $R_V(n, T)$ for SLG and BLG, respectively; the same color coding for the $R_V$ scale. The black curves indicate zero $R_V$. For each $n$ away from the CNP, there is a wide range of $T$ over which $R_V$ is negative. All measurements presented in this work for BLG were taken with zero displacement between the graphene layers (*18*).

The same anomalous vicinity response could also be observed if we followed the recipe of (*15*) and used the current $I$ to increase the electron temperature. In this case, $V_V$ changed its sign as a function of $I$ from positive to negative to positive again, reproducing the behavior of $R_V$ with increasing $T$ of the cryostat (*fig. S10*). Comparing *figs. S8* and *S10*, it is clear that the vicinity geometry strongly favors the observation of hydrodynamics effects so that the measured vicinity voltage changed its sign whereas in the standard geometry the same viscosity led only to relatively small changes in $dV/dI$. We also found that the magnitude of negative $R_V$ decayed rapidly with $L$ (*fig. S11*), in agreement with the finite size of electron whirlpools (see below).

It is important to mention that negative resistances can in principle arise from other effects such single-electron ballistic transport ($\ell_{ee} \gg \ell$) or quantum interference (*18,20,24*). The latter contribution is easily ruled out because quantum corrections rapidly wash out at $T > 20$ K and have a random sign that rapidly oscillates as a function of magnetic field. Also, our numerical simulations using the Landauer-Büttiker formalism and the realistic device geometry showed that no negative resistance could be expected for the vicinity configuration in zero magnetic field (*19,21*). Nonetheless, we carefully considered any 'accidental spillover' of single-electron ballistic effects into the vicinity geometry from the point of view of experiment. The dependences of the negative vicinity signal on $T$, $n$, $I$ and the device geometry allowed us to unambiguously rule out any such contribution as discussed in Supplementary Material. For example, the single-electron ballistic phenomena should become more pronounced for longer $\ell$ (that is, with decreasing $T$ or the electron temperature and with increasing $n$), in stark contrast to the nonmonotonic behavior of $V_V$.

Now we turn to theory and show that negative $R_V$ arises naturally from whirlpools that appear in a viscous Fermi liquid near current-injecting contacts. As discussed in (*18*), electron transport for sufficiently short $\ell_{ee}$ can be described by the hydrodynamic equations

$$\nabla \cdot \boldsymbol{J}(\boldsymbol{r}) = 0 \tag{1}$$

and

$$\frac{\sigma_0}{e} \nabla \phi(\boldsymbol{r}) + D_\nu^2 \nabla^2 \boldsymbol{J}(\boldsymbol{r}) - \boldsymbol{J}(\boldsymbol{r}) = 0 \tag{2}$$

where $J(r) = nv(r)$ is the (linearized) particle current density, and $\phi(r)$ is the electric potential in the 2D plane. If $D_\nu \to 0$, Eq. (2) yields Ohm's law, $-eJ(r) = \sigma_0 E(r)$ with a Drude-like conductivity, $\sigma_0 \equiv ne^2\tau/m$ where $-e$ and $m$ are the electron charge and the effective mass, respectively. The three terms in Eq. (2) describe: the electric force generated by the steady-state charge distribution in response to applied current $I$, the viscous force (*1,2*), and friction due to momentum non-conserving processes parametrized by the scattering time $\tau(n,T)$.

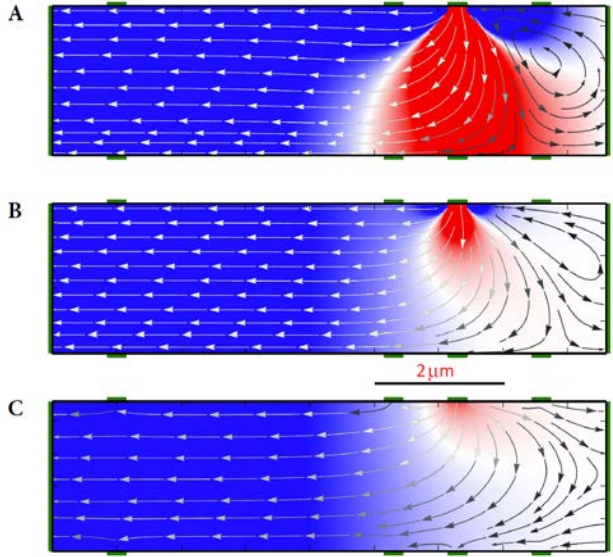

**Fig. 3. Whirlpools in electron flow.** (**A-C**) Calculated $J(r)$ and $\phi(r)$ for the geometry such as that in Fig. 1C, with the green bars indicating voltage contacts. The color scale is for $\phi$ (red to blue corresponds to $\pm I/\sigma_0$). $D_\nu = 2.3$, 0.7 and 0 µm for panels A to C, respectively. Vortices are seen in the top right corners of A and B where the current flow is in the direction opposite to that in (C) that shows the case of zero viscosity. In each panel, the current streamlines also change color from white to black indicating that the current density $|J(r)|$ is lower to the right of the injecting contact.

Equations (1,2) can be solved numerically as detailed in (*18*), and Fig. 3 shows examples of spatial distributions of $\phi(r)$ and $J(r)$. One can clearly see that, for experimentally relevant values of $D_\nu$, a vortex appears in the vicinity of the current-injecting contact. This is accompanied by the sign reversal of $\phi(r)$ at the vicinity contact on the right of the injector, which is positive in Fig. 3C (no viscosity) but becomes negative in Figs. 3A,B. Our calculations for this geometry reveal that $R_V$ is negative for $D_\nu \gtrsim 0.4$ µm (*18*). Because both $\tau$ and $\nu$ decrease with increasing $T$, $D_\nu$ also decreases, and stray currents start to dominate the vicinity response at high $T$. This explains why $R_V$ in Figs. 1-2 becomes positive close to room $T$, even though our hydrodynamic description has no high-temperature cutoff. Note that despite positive $R_V$ the viscous contribution remains quite significant near room $T$ (Fig. 1D, *fig. S12*). On the other hand, at low $T$ the electron system approaches the Knudsen regime and our hydrodynamic description becomes inapplicable because $\ell_{ee} \sim \ell$ (*18*). In the latter regime, the whirlpools should disappear and $R_V$ become positive, in agreement with the experiment and our numerical simulations based on the Landauer-Büttiker formalism.

The numerical results in Fig. 3 can be understood if we rewrite Eqs. (1,2) as

$$\boldsymbol{J}(\boldsymbol{r}) = \frac{\sigma_0}{e}\nabla\phi(\boldsymbol{r}) - nD_v^2 \nabla \times \boldsymbol{\omega}(\boldsymbol{r}),\qquad(3)$$

where $\boldsymbol{\omega}(\boldsymbol{r}) \equiv n^{-1}\nabla \times \boldsymbol{J}(\boldsymbol{r}) = \omega(\boldsymbol{r})\hat{\boldsymbol{z}}$ is the vorticity (2). Taking the curl of Eq. (3), the vorticity satisfies the equation $\omega(\boldsymbol{r}) = D_v^2 \nabla^2 \omega(\boldsymbol{r})$ where $D_v$ plays the role of a diffusion constant. Current $I$ injects vorticity at the source contact, which then exponentially decays over the length scale $D_v$. For $L$ =1 µm, $\nu = 0.1$ m² s⁻¹ and $\tau = 1.5$ ps (18), we find $D_v \approx 0.4$ µm, in qualitative agreement with the measurements in *fig. S11*.

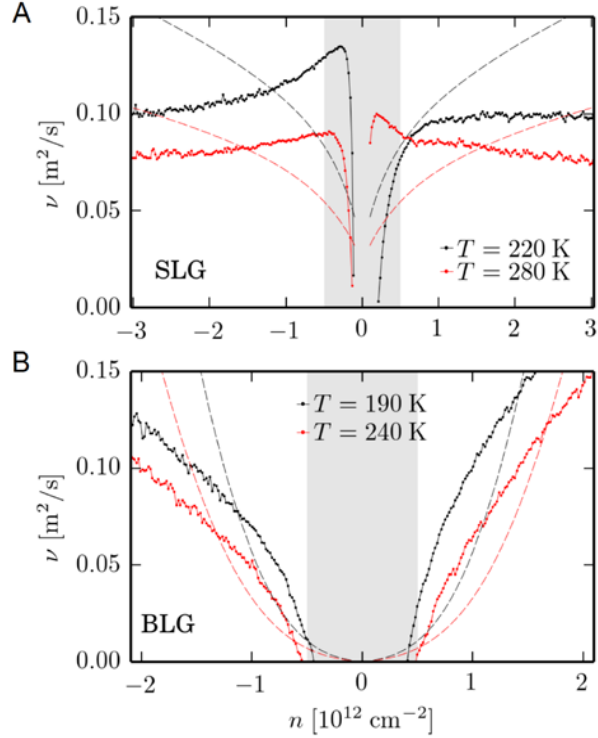

**Fig. 4. Viscosity of the Fermi liquids in graphene.** (**A,B**) Solid curves: $\nu$ extracted from the experiment for SLG and BLG, respectively. Dashed: Calculations based on many-body diagrammatic perturbation theory (no fitting parameters). The grey-shaded areas indicate regions around the CNP where our hydrodynamic model is not applicable (*18*).

Finally, we combine the measurements of $R_V$ and $\rho_{xx}$ with the solution of Eqs. (1,2) in Fig. 3 to extract the kinematic viscosity for SLG and BLG. Because the observed Gurzhi effect in $\rho_{xx}$ is small at low currents (*fig. S6*), we can use $\rho_{xx} = 1/\sigma_0 = m/(ne^2\tau)$ to determine $\tau(n,T)$ (*18*). Furthermore, for the experimentally relevant values of $D_v$, we find that $R_V$ is a quadratic function of $D_v$

$$R_V = (b + a\, D_v^2)\sigma_0^{-1} \qquad(4)$$

where $a$ and $b$ are numerical coefficients dependent only on the measurement geometry and boundary conditions and $b$ describes the contribution from stray currents (*fig. S13*). For the specific device in Fig. 3, we determine $a = -0.29$ µm⁻² and $b = 0.056$, and this allows us to estimate $D_v(n,T)$ from measurements of $R_V$. For the known $\tau$ and $D_v$, we find $\nu(n,T) = D_v^2/\tau$. Applicability limits of this analysis are discussed in Supplementary Material, and the results are plotted in Fig. 4 for one of our devices. It shows that, at carrier concentrations $\sim 10^{12}$ cm⁻² the Fermi liquids in both SLG and BLG are highly viscous with $\nu \approx 0.1$ m$^2$s$^{-1}$. For the sake of comparison, liquid honey has typical viscosities of $\approx 0.002$-$0.005$ m$^2$s$^{-1}$.

Fig. 4 also plots results of fully-independent microscopic calculations of $\nu(n,T)$, which were carried out by extending the many-body theory of ref. (*25*) to the case of 2D electron liquids hosted by SLG and BLG. Within the range of applicability of our analysis in Fig. 4 ($n \sim 10^{12}$ cm$^{-2}$), the agreement in absolute values of the electron viscosity is remarkable, especially taking into account that no fitting parameters were used in the calculations. Because the strong inequality $\ell \gg \ell_{ee}$ required by the hydrodynamic theory cannot be reached even for graphene, it would be unreasonable to expect better agreement (*18*). In addition, our analysis does not apply near the CNP because the theory neglects contributions from thermally-excited carriers, spatial charge inhomogeneity and coupling between charge and energy flows, which can play a substantial role at low doping (*16,18*). Further work is needed to understand electron hydrodynamics in the intermediate regime $\ell \gtrsim \ell_{ee}$ and, for example, explain ballistic transport ($\ell > W$) in graphene at high $T$ in terms of suitably modified hydrodynamic theory. Indeed, the naïve single-particle description that is routinely used for graphene's ballistic phenomena even above 200 K (*19,21*) cannot be justified and needs to be explained in terms of electron-liquid jets. As for experiment, the highly viscous Fermi liquids in graphene and their accessibility offer a tantalizing prospect of using various scanning probes for visualization and further understanding of electron hydrodynamics.

**Acknowledgments:** This work was supported by the European Research Council, the Royal Society, Lloyd's Register Foundation, the Graphene Flagship and the Italian Ministry of Education, University and Research through the program Progetti Premiali 2012 (project ABNANOTECH). D.A.B. and I.V.G. acknowledges Marie Curie program grant SPINOGRAPH.

## Supplementary Material

### #1 Device fabrication

Our devices were made from single- and bi-layer graphene encapsulated between relatively thick (~50 nm) crystals of hexagonal boron nitride (hBN). The crystals' transfers were carried out using the dry-peel technique described previously (*20,26*). The heterostructures were assembled on top of an oxidized Si wafer (300 nm of $SiO_2$) which served as a back gate, and then annealed at 300°C in Ar-$H_2$ atmosphere for 3 hours. After this, a PMMA mask was fabricated on top of the hBN-graphene-hBN stack by electron-beam lithography. This mask was used to define contact areas to graphene, which was done by dry etching with fast selective removal of hBN (*27*). Metallic contacts (usually, 5 nm of Ta followed by 50 nm Nb) were then deposited onto exposed graphene edges that were a few nm wide. Such quasi-one-dimensional contacts to graphene (*27*) had notably lower contact resistance than those reported previously without the use of selective hBN etching (*20*). As the next step, another round of electron-beam lithography was used to prepare a thin metallic mask ($\approx$ 40 nm Al) which defined a multiterminal Hall bar. Subsequent plasma etching translated the shape of the metallic mask into encapsulated graphene (see *figs*. *S1A-B* and Fig. 1C of the main text). The Al mask could also serve as a top gate, in which case Al was wet-etched near the Nb/Ta leads to remove the electrical contact to graphene. All our bilayer graphene (BLG) devices were prepared with such a top gate, which allowed us to control not only the carrier concentration but also the displacement field between the two layers. Also, for single-layer graphene (SLG) we usually (but not always) made both top and bottom gates for the sake of fabrication procedures, even though the two gates fulfilled essentially the same function.

The studied Hall bars were 1.5 to 4 μm in width $W$ and up to 20 μm in length (larger $W$ were avoided as we previously found them to suffer from charge inhomogeneity induced by contamination bubbles and associated strain; ref. *28*). The devices were carefully characterized and, in addition to Fig. 1D of the main text, an example of typical behavior of $\rho_{xx}(n)$ is shown in *fig. S1C*. All the studied devices, independently of their width or length, were found to exhibit negative vicinity resistance over the described range of temperatures below room $T$ and over a wide range of $n \sim 10^{12}$ cm$^{-2}$. *Fig. S1D* shows another example of this behavior, which is rather similar to that in Fig. 1E of the main text.

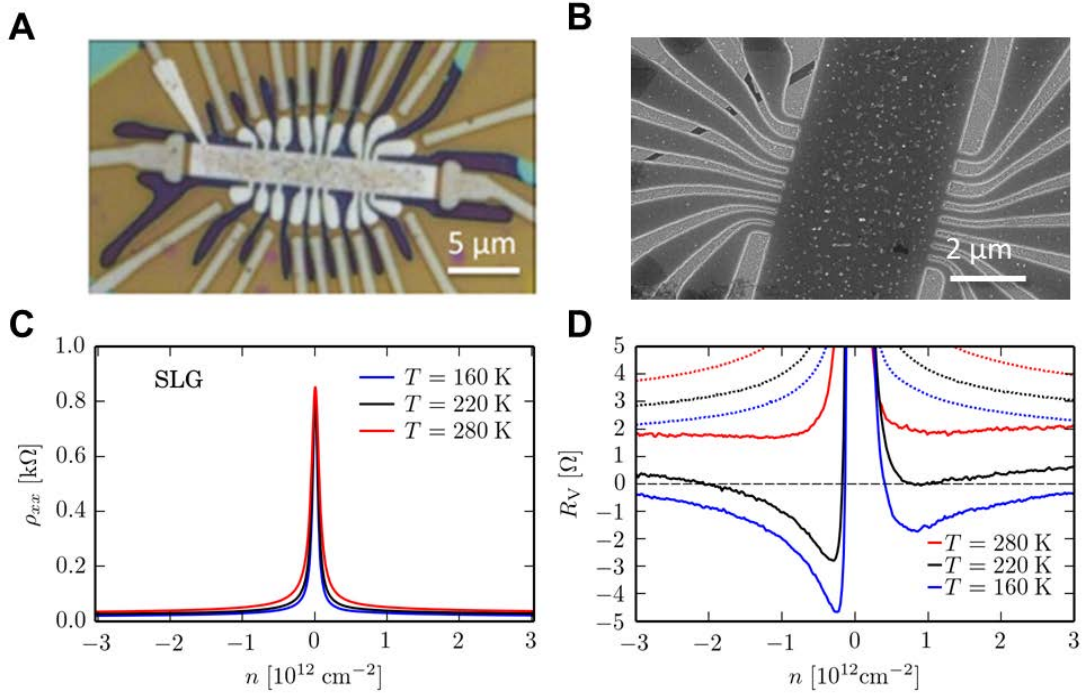

**Fig. S1. Further examples of the studied graphene devices and their behavior.** (**A**) Optical micrograph for an encapsulated SLG device. The bright white area is the top gate and the graphene Hall bar repeats its shape. Numerous metallic leads terminated with quasi-one-dimensional contacts to graphene are seen in a duller white color. Other colors on the photo appeared due to different etching depths of the hBN-graphene-hBN stack. (**B**) Electron micrograph of yet another SLG device. Resistivity (**C**) and vicinity resistance (**D**) for the device shown in (A). For resistivity measurements, we always used voltage probes separated by a distance larger than the main channel width. In (C), voltage probes were 8 µm away from each other. The vicinity probe used in (D) was 1 µm away from the current injecting lead. Positive and negative sign of $n$ correspond to gate-induced electrons and holes, respectively. The dashed curves in (D) show the expected 'classical' contribution $b\rho_{xx}$ which arises due to stray currents. For this particular device, we find $b \approx 0.1$ using numerical simulations of the device geometry as described in the main text and the supplementary section on numerical simulation.

#2 Mobility and scattering times

Our longitudinal measurements allowed us to determine $\mu(n,T)$ and $\tau(n,T)$ using the standard relation, $\sigma_{xx} = ne\mu = ne^2\tau/m$. Results are shown in *fig. S2* for both SLG and BLG. The plotted behavior is universal, that is, it changes little between different devices because, for the shown $T$ range of interest, electron transport was limited by electron-phonon scattering. One can see that away from the charge neutrality point (CNP), $\tau$ depends weakly on $n$ for both SLG and BLG. Typical times are of about 1-2 ps. As for $\mu(n,T)$, its behavior as a function of $n$ is notably different in the two graphene systems because of different energy dependences of their effective masses. For BLG, which has a nearly parabolic spectrum, we can for simplicity use the constant $m = 0.03m_0$ where $m_0$ is the free electron mass.

This yields that $\mu$ is simply proportional to $\tau$. For SLG, the effective (or cyclotron) mass is given by $m \propto \sqrt{n}$, leading to $\mu$ varying approximately as $n^{-1/2}$ (*fig. S2B*).

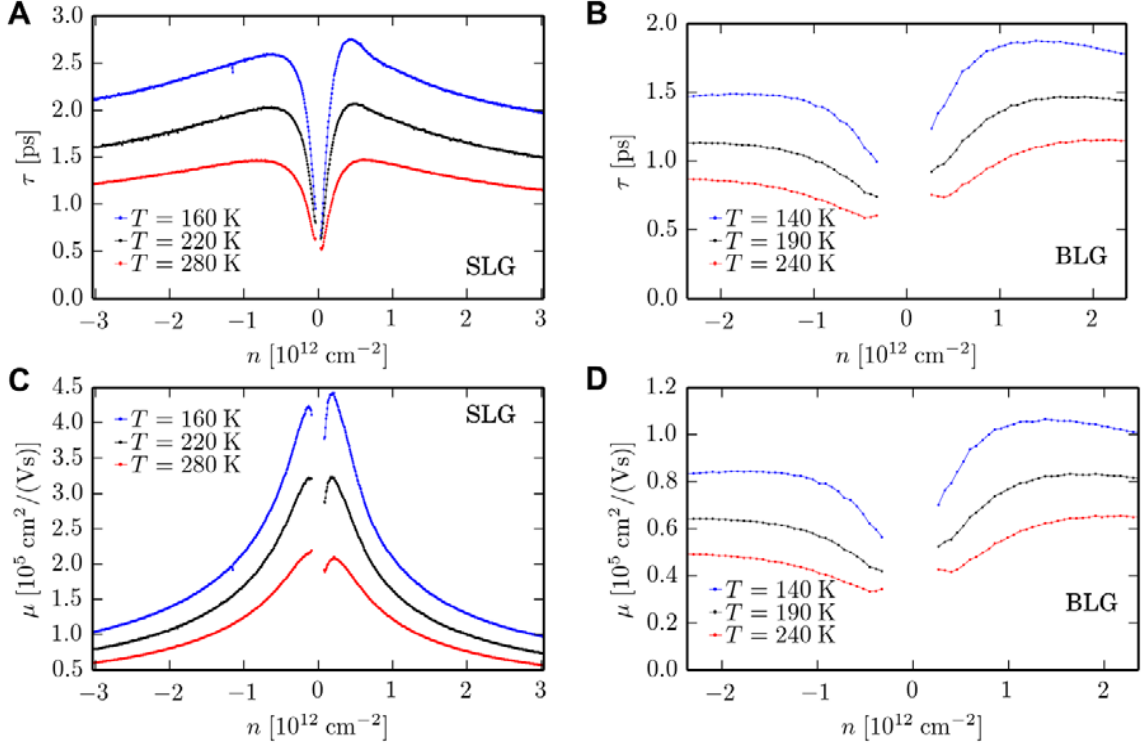

**Fig. S2. Phonon-limited transport in graphene.** (**A,B**) Mean free times and mobilities in encapsulated SLG, respectively. (**C,D**) Same for encapsulated BLG. The plots are for the $T$ range in which hydrodynamics effects were found strongest.

#### #3 Microscopic calculations of the electron-electron mean free path
In this Section we briefly summarize the results of many-body diagrammatic perturbation theory calculations of the electron-electron scattering length $\ell_{ee}$. Results in this Section refer to SLG, in the region of parameter space relevant for our experiments.

We calculated $\ell_{ee} = v_F \tau_{ee}$ from the imaginary part of the retarded quasiparticle self-energy $\Sigma_\lambda(k,\omega)$, evaluated at the Fermi surface. For an electron doped system we find $\hbar/\tau_{ee} = -2\,\Im m[\Sigma_{\lambda=+1}(k_F,0)]$. Here, $v_F$ is the (bare) Fermi velocity (which is equal to the Dirac velocity $v_D$ in SLG and $\hbar k_F/m$ in BLG), $\lambda = \pm 1$ is a conduction/valence band index, and $\tau_{ee}$ is the quasiparticle lifetime due to e-e scattering (*14*). The quantity $\Sigma_\lambda(k,\omega)$ can be calculated by using the $G_0W$ approximation (*23*) with a dynamically screened interaction $W_{k,\omega}$ evaluated at the level of the random phase approximation (RPA) (*14*).

In the case of SLG, the imaginary part of the quasiparticle self-energy is given by the following expression (23)

$$\Im m[\Sigma_\lambda(k,\omega)] = -\sum_{\lambda'=\pm 1}\int \frac{d^2\boldsymbol{q}}{(2\pi)^2}\Im m\left[W_{\boldsymbol{q},\omega-\xi_{\lambda',\boldsymbol{k}+\boldsymbol{q}}}\right] F_{\lambda\lambda'}[n_{\rm B}\left(\hbar\omega-\xi_{\lambda',\boldsymbol{k}+\boldsymbol{q}}\right)+n_{\rm F}(-\xi_{\lambda',\boldsymbol{k}+\boldsymbol{q}})]$$

where $F_{\lambda\lambda'} = [1+\lambda\lambda'\cos(\theta_{\boldsymbol{k},\boldsymbol{k}+\boldsymbol{q}})]/2$ is the chirality factor, $\xi_{\lambda,\boldsymbol{k}} = \lambda\hbar v_{\rm F}k - \mu$ is the Dirac band energy measured with respect to the chemical potential $\mu$, $W_{\boldsymbol{q},\omega} = v_q/\varepsilon(q,\omega) \equiv v_q/[1-v_q\chi_0(q,\omega)]$ the RPA dynamically screened interaction, and $n_{\rm B/F}(x) \equiv 1/[\exp(\beta x)\mp 1]$ are the Bose/Fermi statistical factors with $\beta = 1/(k_{\rm B}T)$. In the above expressions, $v_q$ is a suitably-chosen effective Coulomb interaction (see below), and $\chi_0(q,\omega)$ is the polarization function of a non-interacting 2D massless Dirac fermion system at a finite temperature and carrier density (29). More details can be found, for example, in Refs. (22,23).

In our calculations we have also estimated the impact of the 'environment' such as i) nearby metal gates (by modeling them as perfect conductors), ii) the uniaxial anisotropy of dielectric hBN, and iii) thin-film effects. The bare Coulomb potential $2\pi e^2/q$ is strongly modified by these three factors. The effective Coulomb interaction $v_q$ can be written in the form $2\pi e^2 \mathcal{G}(qd, qd')/q$ where the explicit functional dependence of the form factor $\mathcal{G}(x,y)$ on its variables $x$ and $y$ is rather cumbersome and will be reported elsewhere. The form factor depends on the thickness $d$ and $d'$ of the hBN slab below and above graphene, respectively. It also depends on the static values of the in-plane $\epsilon_x(\omega)$ and out-of-plane $\epsilon_z(\omega)$ components of the hBN permittivity tensor: see, for example, Ref. (30). Numerical results for $\ell_{\rm ee}$ in encapsulated SLG are shown in *fig. S3*.

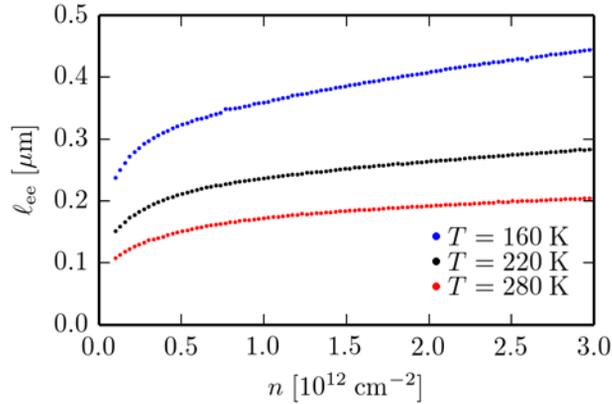

**Fig. S3. Numerical results for the e-e mean free path $\ell_{\rm ee}$ in our encapsulated SLG devices.** Results are shown as a function of carrier density $n$ and for three $T$. For this particular calculation, we used $d = 80$ nm and $d' = 70$ nm and took into account the top metal gate simulating the device shown in *fig. S1A*. We have also checked that metal gates at such distances play little role (the presence of the gate changed $\ell_{\rm ee}$ typically by less than 5% with respect to the ungated case), in agreement with the fact that the SLG devices with and without top gates exhibited the $R_{\rm V}$ behavior indistinguishable within variations between different contacts.

Besides determining the region of parameter space where the hydrodynamics theory can be applied, the frequency of electron-electron collisions also determines the numerical value of the electron liquid viscosity. The usual estimate for the value of the kinematic viscosity of a classical liquid is $\nu \sim v\ell_{\text{coll}}$ (*31*), where $v$ is a characteristic velocity (e.g. the thermal velocity for classical liquids) of particles and $\ell_{\text{coll}}$ is the mean free path for inter-particle collisions.

Microscopic calculations for SLG yield (*25*):
$$\nu = \frac{1}{4} v_F \tilde{\ell}_{\text{ee}} \tag{S1}$$
where $\tilde{\ell}_{\text{ee}}$ is a characteristic length associated with electron-electron scattering, which is of the same order of magnitude as $\ell_{\text{ee}}$ in the explored range of parameters. Their ratio $\tilde{\ell}_{\text{ee}}/\ell_{\text{ee}}$ is shown in *fig. S4*. Eq. (*S1*) is consistent with the above estimate for classical fluids. From Eq. (*S1*) we also find that the viscosity diffusion length $D_\nu = \sqrt{\nu\tau}$, which determines the size of electron whirlpools, is equal to $D_\nu = \sqrt{\tilde{\ell}_{\text{ee}}\ell}/2$ and, therefore, depends on both electron-electron collisions and momentum-non-conserving collisions.

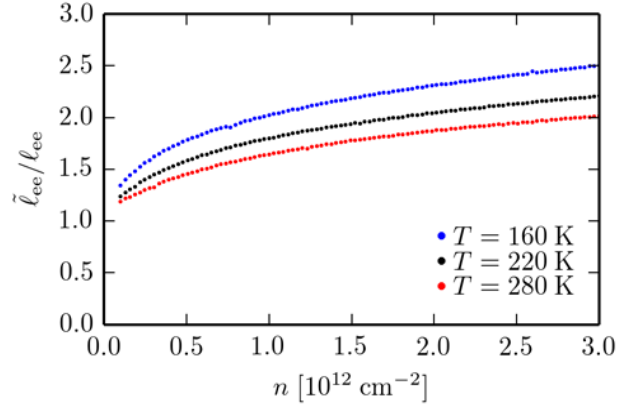

**Fig. S4. Comparison between $\ell_{\text{ee}}$ and $\tilde{\ell}_{\text{ee}}$.** Numerical results for the ratio $\tilde{\ell}_{\text{ee}}/\ell_{\text{ee}}$ in the same range of densities and for the same temperatures as in *fig. S3*.

**#4 On pseudo-relativistic and pressure terms in the Navier-Stokes equation**

Because of the pseudo-relativistic nature of transport in SLG, the Navier-Stokes equation for the two-dimensional electron liquid in SLG contains a number of pseudo-relativistic terms (*32*). Such terms have not been considered in Eq. (2) of the main text. It is possible to demonstrate that, if one considers only linear deviations from a situation of uniform and static equilibrium ($n(\mathbf{r},t) = n$ and $\mathbf{v}(\mathbf{r},t) = \mathbf{0}$), the only pseudo-relativistic correction that survives is the appearance of the effective (cyclotron) mass $m = \hbar k_F/v_D$ in the definition of the Drude-like conductivity $\sigma_0$ for the case of SLG.

In deriving Eq. (2) of the main text we have also neglected a term arising from the pressure $P$ of the electron liquid, i.e. $-\nabla P(\mathbf{r},t)$. Here we show that, for a gated structure like the one used in our experiments, this term is simply proportional to the electric field and its only effect is to give a small correction to the capacitance between the graphene sheet and the top and bottom gates. In a gated structure, the electric potential and carrier density can be related by the so-called local capacitance approximation (LCA) (*32*), i.e. $\phi(\mathbf{r},t) = -en(\mathbf{r},t)/C$ where $C$ is the capacitance per unit area. For a double gated device, $C = \epsilon(d+d')/(4\pi d\, d')$ where $d$ and $d'$ are the distances between graphene and the bottom and top gates, respectively, and $\epsilon = \sqrt{\epsilon_x(0)\epsilon_z(0)} \approx 4.4$ is the static dielectric constant of bulk hBN (we neglect the thin-film effects discussed in the previous Section). The quantities $\epsilon_x(\omega)$ and $\epsilon_z(\omega)$ have been introduced in the previous Section.

Using the LCA and the local density approximation $\nabla P(\mathbf{r},t) \approx (\partial P_{\text{hom}}/\partial n)\nabla n(\mathbf{r},t)$, we can estimate the sum of the electric force and the force due to pressure as following

$$-\left(\frac{e^2 n}{C} + \frac{\partial P_{\text{hom}}}{\partial n}\right)\nabla n(\mathbf{r},t) \tag{S2}$$

where $P_{\text{hom}}$ is the pressure of the homogeneous 2D electron liquid in SLG or BLG. Evaluating the two terms inside the round brackets at the equilibrium density $n$ and approximating $\partial P_{\text{hom}}/\partial n$ with its zero-temperature non-interacting value, i.e., $\partial P_{\text{hom}}/\partial n \approx \xi E_{\text{F}}$ where $\xi = 1/2$ ($\xi = 1$) for SLG (BLG), we can show that the ratio between the pressure term and the potential term is

$$\frac{\partial P_{\text{hom}}/\partial n}{e^2 n/C} \approx \frac{d+d'}{4dd'\ \xi k_{\text{TF}}} \tag{S3}$$

where $k_{\text{TF}}$ is the Thomas-Fermi screening wave number. For encapsulated SLG, $k_{\text{TF}} = 4\,\alpha_{\text{ee}} k_{\text{F}} \approx (2.9\text{ nm})^{-1}$, where $\alpha_{\text{ee}} = e^2/(\hbar v_{\text{D}}\epsilon) \approx 0.5$ is the so-called graphene fine structure constant (*29*). Using a carrier density of $10^{12}\text{ cm}^{-2}$, $d = 80$ nm and $d' = 80$ nm, we find that the ratio in Eq. (*S3*) is much smaller than unity. The pressure term can be safely neglected. For an encapsulated BLG sheet $k_{\text{TF}} = 2e^2 m/(\hbar^2 \epsilon) \approx (3.8\text{ nm})^{-1}$, irrespective of density (*14*). Therefore, the ratio (*S3*) is also negligible in this case.

#### #5 Smallness of the Reynolds number
The validity of the linearized Navier-Stokes equation (Eq. (2) of the main text) relies on the smallness of the Reynolds number (*1*), a dimensionless parameter that depends on the sample geometry and controls the smallness of the nonlinear term $[\mathbf{v}(\mathbf{r},t) \cdot \nabla]\mathbf{v}(\mathbf{r},t)$ in the convective derivative in the full Navier-Stokes equation with respect to the viscous term. In our case

$$\left|\frac{[\mathbf{v}(\mathbf{r},t)\cdot\nabla]\mathbf{v}(\mathbf{r},t)}{\nu\nabla^2 \mathbf{v}(\mathbf{r},t)}\right| \approx \frac{|\mathbf{v}|W}{\nu} = \frac{I}{en\nu} \equiv \mathcal{R} \tag{S4}.$$

For a typical probing current $I = 10^{-7}$ A, $W = 1$ μm and $n = 10^{12}\text{ cm}^{-2}$, we estimate $|\mathbf{v}| \sim I/(enW) \approx 10^4$ cm/s. The corresponding value of the Reynolds number is $\mathcal{R} \sim 10^{-3} \ll 1$ if using $\nu \sim 10^3$ cm$^2$/s found theoretically (*25*) and in the experiment (Fig. 4 of the main text). Our linearized approximation is therefore fully justified.

#6 On the boundary conditions for solid-state hydrodynamic equations

The hydrodynamic equations need to be accompanied by appropriate boundary conditions (BCs). If viscosity is negligible, the current is proportional to the gradient of the potential. In this case it is sufficient to solve the Laplace equation for the potential to obtain both potential and current spatial patterns. The BCs that the potential must obey at the boundaries of the sample are of two types. In regions of the boundary where no electrical contacts are present, the normal component of the current (that is the normal derivative of the potential in the non-viscous regime) must be zero. In the regions of the boundaries where an electrical contact is present, the potential immediately inside the sample must be equal to the electric potential of the contact. Since the sample is current biased, we fix the total current flowing from each contact instead of fixing the value of the potential at each contact. It can be shown using standard theorems on the Laplace equation that these BCs (Neumann outside the contacts and Dirichlet at the contacts) uniquely determine the solution of the problem.

In the general case of a viscous flow, Eq. (2) of the main text requires additional BCs on the tangential component of the current. Generally, edges exert friction on the 2D electron liquid. The corresponding force (per unit length) is given by (*1*)

$$F_t = \epsilon_{ij} \hat{n}_i \sigma'_{jk} \hat{n}_k \qquad (S5).$$

In Eq. (*S5*), $\sigma'_{jk}$ is the 2D viscous stress tensor, i.e. $\sigma'_{jk} = \eta(\partial_j v_k + \partial_k v_j - \delta_{ij}\partial_l v_l)$. In writing the previous expression for $\sigma'_{jk}$ we have set to zero the diagonal contribution that is proportional to the so-called bulk viscosity and negligible (*25*).

The frictional force is in general a function of the tangential velocity $v_t = \epsilon_{ij} \hat{n}_i v_j$. For small velocities the force is simply proportional to the velocity leading to the BC

$$\epsilon_{ij} \hat{n}_i \hat{n}_k (\partial_j v_k + \partial_k v_j) = \epsilon_{ij} \hat{n}_i v_j / l_\text{b} \qquad (S6)$$

where $l_\text{b}$ is a characteristic length scale associated with boundary scattering. If this length is very small, Eq. (*S6*) reduces to the standard "no-slip" boundary condition commonly used in the description of classical liquids (*1*).

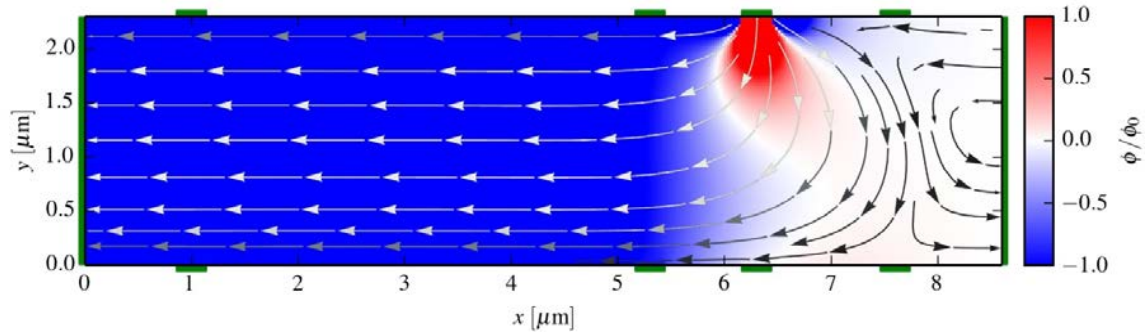

**Fig. S5. Influence of boundary conditions.** Calculated current density $J(r)$ and electric potential $\phi(r)$ for the same geometry and the same $D_v = 0.7$ μm as in Fig. 3B of the main text. The difference is no-slip boundary conditions ($l_\text{b}=0$) used in this figure whereas $l_\text{b}=\infty$ in the main text.

We do not know precisely the value of $l_\mathrm{b}$ but we know that the combined effect of friction and viscosity arising at the boundaries can lead to an anomalous temperature dependence of the longitudinal resistivity $\rho_{xx}$ which is known as the Gurzhi effect (*7*). As discussed in one of the following Sections, our experimental data for $\rho_{xx}$ exhibit a monotonic behavior as a function of $T$, up to our highest temperature and for all carrier densities. This behavior suggests that the Gurzhi effect is small. For this reason, we can use to a good approximation the following BCs (*12*)

$$\epsilon_{ij}\hat{n}_i\hat{n}_k(\partial_j v_k + \partial_k v_j) = 0 \tag{S7}$$

which assumes that $l_\mathrm{b}$ is larger than the characteristics length scales of the problem, $D_v$ (vorticity diffusion length) and $W$ (width of our multiterminal devices). Physically, Eq. (*S7*) corresponds to a vanishing tangential force acting on a moving liquid (*1*). At high current densities, however, the friction from the boundaries can be enhanced with respect to the simple linear model in Eq. (*S6*). In this case the Gurzhi effect can be observed in the differential resistance (see below).

In *fig. S5* we show that different values of $l_\mathrm{b}$ have little impact on the formation of whirlpools near current injecting contacts. The reader is urged to compare *fig. S5* (no-slip boundary conditions) with Fig. 3B in the main text, which was obtained using the free-surface BCs (Eq. *S7*).

**#7 Applicability limits for hydrodynamic description of electron transport in doped graphene**
The focus of our report is on the doped regime because the situation near the CNP is severely complicated by the presence of thermally excited quasiparticles, electron-hole puddles (*33*) and the large electron wavelength. In addition, thermoelectric effects (energy flow) are also expected to play a significant role near the CNP, although they appear only in the second order with respect to applied current $I$ in zero magnetic field (see, for example, ref. *34*).

Under realistic experimental conditions, one important limit is set by charge inhomogeneity that impacts the viscosity analysis presented in Fig. 4 of the main text. Indeed, Eq. (4) assumes that $\sigma_0 = 1/\rho_{xx}$ is constant whereas the inhomogeneity locally modifies conductivity and stray currents. The electron-hole asymmetry seen in the experimental plots for $R_\mathrm{V}$ and the associated asymmetry in Fig. 4 of the main text are not expected in theory, and this provides a qualitative indication of the best accuracy one can expect for the extracted values of $\nu$.

Our hydrodynamic theory suggests no high-$T$ cutoff, at least up to temperatures at which optical phonon scattering starts playing a role. In fact, the theory smoothly converges with the standard Drude theory as viscosity tends to zero upon increasing $T$. However, there is a clear high-$T$ cutoff on $R_\mathrm{V}$ being negative. It is simply dictated by the two competing terms in Eq. (4) of the main text, which are due to stray currents and viscous flow. After subtracting the stray-current contribution from the measured vicinity resistance (see below), we find that the hydrodynamic term smoothly extend to high $T$ over the entire temperature range without any sign of cutoff.

On the other hand, the essential condition of electron hydrodynamics ($\ell_{ee} \ll \ell$) certainly fails at temperatures below 50 K where the phase breaking length in graphene (which is smaller than $\ell_{ee}$) is known to reach a micrometer scale (see, e.g., ref. *35*) and electron transport can be described in terms of single-particle ballistics (billiard-ball model). Our hydrodynamic theory does not capture the crossover ($\ell_{ee} \sim \ell$) into this single-particle regime, and it remains to be investigated how strong the above inequality condition should be to allow the hydrodynamic description.

#8 Absence of the Gurzhi effect in longitudinal resistivity

Resistivity of an electron liquid is determined by interplay between bulk scattering (charged impurities, lattice vibrations, crystal defects, etc.), collisions at sample boundaries and e-e scattering (*7,15*). Bulk scattering normally increases with $T$. On the other hand, a combined effect of boundary and electron-electron scattering results in a contribution to $\rho_{xx}$ which increases with $T$ if $\ell_{ee} \gg W$ (Knudsen regime) but decreases if the electron system enters the viscous flow regime, $\ell_{ee} \ll W$. The transition between the two limits may result in a non-monotonic temperature dependence of $\rho_{xx}$. This phenomenon is referred to as the Gurzhi effect (*7,15*). In reality, this effect is severely obscured by various bulk scattering mechanisms and expected to be weak (*15*).

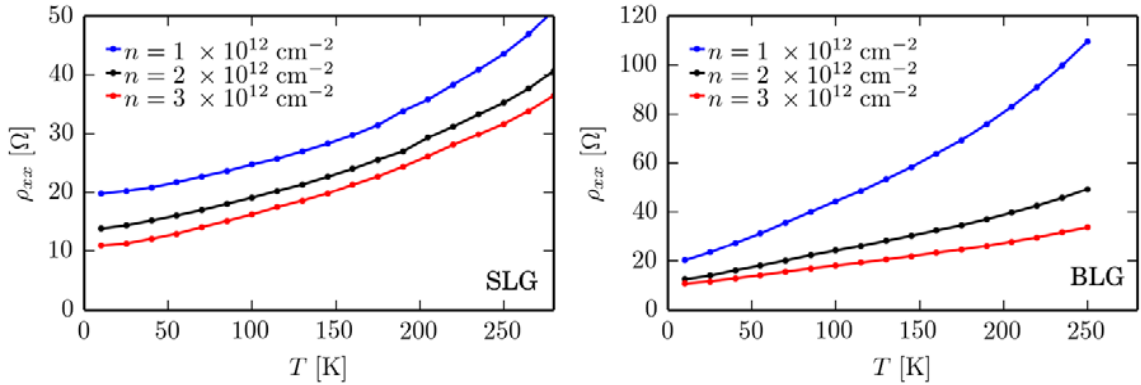

**Fig. S6. Temperature dependence of longitudinal resistivity**. Left and right panels are for SLG and BLG devices, respectively. The *T* dependences are monotonic, although one can notice that the curves slightly bulge around 100 K, which we attribute to a small hydrodynamics contribution related to the Gurzhi effect, as discussed in the next section.

In *fig. S6*, we show typical measurements of $\rho_{xx}$ as a function of $T$ for our SLG and BLG devices at different carrier concentrations. The behavior of $\rho_{xx}(T)$ is monotonic (no Gurzhi effect) even in the region of parameter space where electron-electron scattering is strong enough to cause the observed sign change in the vicinity geometry. This can be attributed to relative insensitivity of electron flow to boundary scattering in this simplest geometry of measurements as discussed in the preceding section. Neglecting more subtle effects observed in the differential resistance (see the next section), the absence of the Gurzhi effect in $\rho_{xx}(T)$ justifies our choice of (free-surface) BCs described by Eq. (*S7*), in which the force exerted by the boundary on the electron fluid flow is neglected.

Using the BC of Eq. (*S7*), we have solved numerically the linearized steady-state hydrodynamic equations for the longitudinal geometry and the results are plotted in *fig. S7* for SLG and BLG. This figure shows that $\rho_{xx}$ depends only on the phenomenological scattering time $\tau$ in the Navier-Stokes equation (Eq. (2) in the main text) and exhibits little dependence on $D_v$ and, hence, the electron viscosity $\nu$. This is the reason why we can use $\rho_{xx}(n,T)$ to find $\tau(n,T)$ and, more generally, why the previous literature on electron transport in graphene, which completely neglected high electron viscosity, does not require revision if the measurements were carried out in the standard longitudinal geometry.

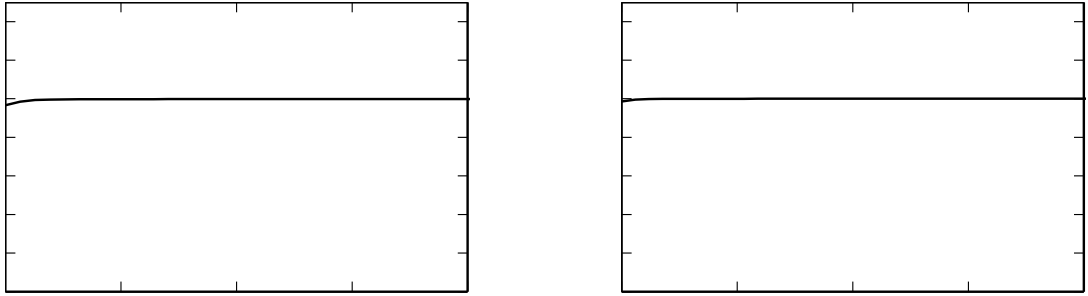

**Fig. S7. Numerical solutions of the linearized hydrodynamic equations in the longitudinal geometry.** In these plots we show the calculated longitudinal resistivity $\rho_{xx}$ as a function of $D_v$ (in µm) for SLG (left) and BLG (right). In solving the hydrodynamic equations we have utilized the free-surface BCs of Eq. (*S7*). From these numerical results, we infer that $\rho_{xx}$ is simply equal to the inverse of the Drude-like conductivity $\sigma_0 \equiv ne^2\tau/m$.

**#9 Gurzhi effect with increasing the electron temperature**

Despite the absence of notable deviations in $\rho_{xx}(T)$ from a monotonic behavior, evidence for the Gurzhi effect could clearly be observed in the longitudinal differential resistance $dV/dI$ measured as a function of a large applied current $I$. The current increased the temperature of the electron system well above the graphene lattice temperature and cryostat's temperature, $T$. Accordingly, these measurements enhanced electron-electron scattering whereas electron-phonon scattering remained relatively weak. Therefore, the $dV/dI$ curves can qualitatively be viewed as changes in $\rho_{xx}$ induced by increasing the electron temperature. Examples of the observed $dV/dI$ as a function of $I$ are shown in *fig. S8*.

At carrier concentrations $|n| > 1 \times 10^{12}$ cm$^{-2}$ we observed rather featureless $dV/dI$ curves up to our highest $I \approx 300$ µA (*fig. S8A*). For smaller $|n|$, the behavior of $dV/dI$ became strongly nonmonotonic, which can be attributed to the Gurzhi effect (*7,15*). We interpret the observed nonlinearity as follows (*15*). At low $T$ and low $I$, electron-electron scattering is weak ($\ell_{ee} \gtrsim W$), and we are in the Knudsen-like regime where the viscosity is determined by scattering at the channel edges. In this regime, resistivity grows with increasing the electron temperature (increasing $I$), similar to the case of classical dilute gases. At higher $I$ ($> 50$ µA), the further increase in the electron temperature pushes the system into the Navier-Stokes regime with $\ell_{ee}$ becoming shorter than $W$. In this case, the flow starts being ruled by internal electron viscosity. The transition between the two regimes is known to lead to a drop in flow resistivity, as first observed by Knudsen for classical gases and, more recently (*15*), reported as the Gurzhi effect for electrons, also using the $dV/dI$ measurements. The $T$ dependence in *fig. S8B* shows that $I \sim 100$ μA heats up the electron system to $\sim 200$ K, which leads to the transition into the Navier-Stokes regime. This is in good agreement with the $T$ range where our hydrodynamic effects were found strongest. Also, note that the Gurzhi effect appeared within the same range of carrier concentrations in which we observed largest negative $R_V$ (compare *fig. S8* with Fig. 2A of the main text and *fig. S9*).

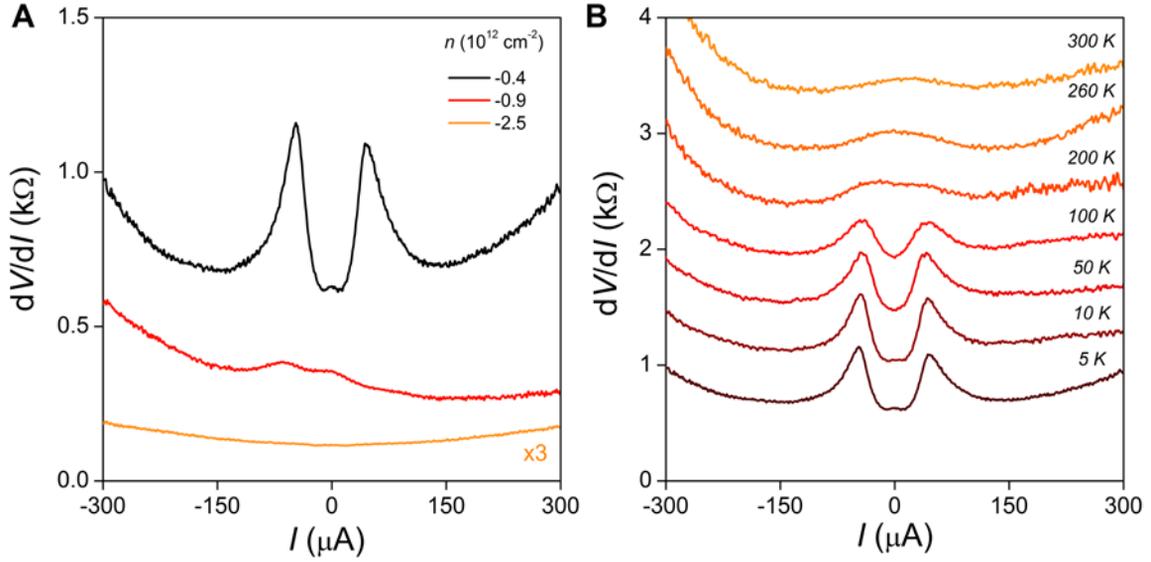

**Fig. S8. Longitudinal differential resistance.** (**A**) Examples of $dV/dI$ as a function of applied current for a SLG device. To measure $dV/dI$, we applied an oscillating current $I + I_{ac}\cos(\omega t)$ along the main channel where $I_{ac}$ is the low-frequency current, much smaller than $I$. The ac voltage drop that appeared at side contacts yielded $dV/dI$. The main channel was 2.5 μm wide, and voltage probes were separated by 8 μm. $T = 5$ K; $I_{ac} = 50$ nA. (**B**) Temperature dependence of the differential resistance in (A) for hole doping with $n = -0.4 \times 10^{12}$ cm$^{-2}$. The curves in (B) are offset for clarity by 300 Ohms each.

#10 Reproducibility of negative vicinity response

To illustrate that the observed whirlpool effects were reproducible for different devices and using different contacts, *fig. S9* shows two more examples of $R_V$ maps. They are for SLG devices with low-$T$ $\mu$ of $\approx 50$ m$^2$ V$^{-1}$ s$^{-1}$ and the distance $L$ to the nearest vicinity probe of $\approx 1$ μm. These maps are rather similar to those shown in Fig. 2 of the main text. Again, we observed large negative vicinity resistance away from the CNP and over a large range of $T$ and $n$.

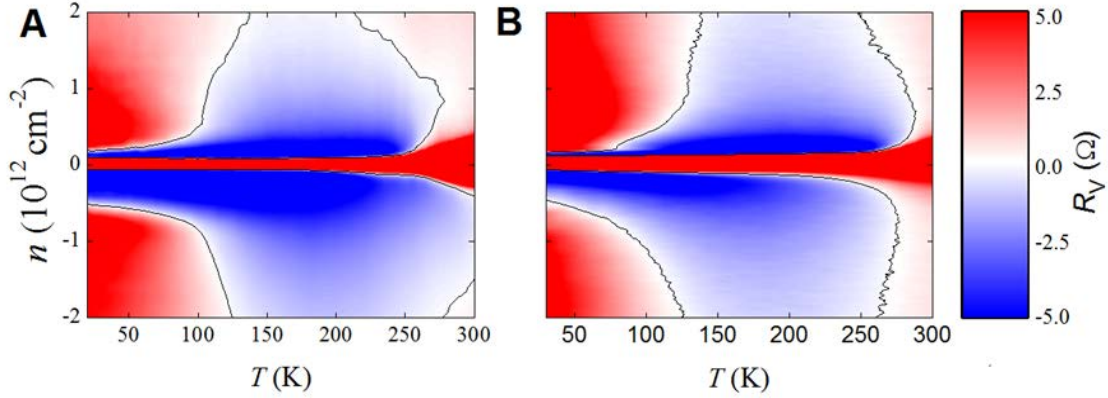

**Fig. S9. Further examples of negative vicinity resistance.** (**A**) and (**B**) are $R_V(n,T)$ maps for two different high-quality SLG devices and the distance between the injection and vicinity contacts of $1$ μm.

**#11 Changes from normal flow to backflow induced by electron heating**

Negative vicinity voltage $V_V$ is attributed to electron whirlpools and expected only in the viscous-flow regime. This requires $\ell_{ee}$ to be smaller than the characteristic scale in the problem, that is, the distance $L$ between the injector and voltage contacts. In addition, to be detectable in transport experiments whirlpools should be sufficiently large in size (large $D_v$) to reach from the injection region to voltage probes. Because $\ell_{ee}$ depends on the electron temperature, the nature of electron flow can be controlled not only by changing the lattice temperature as in the experiments described in the main text but also by heating up the electron system using large dc currents $I$ as discussed in the above section on the Gurzhi effect. We have carried out such measurements of $V_V$ as a function of the electron temperature, and examples of the observed negative vicinity response are shown *fig. S10A*. It plots typical behavior of $V_V$ as a function of $I$ for three characteristic temperatures of the cryostat, *T*. For the case of low *T*, the *I-V* curve exhibits a positive slope at small $I$ which corresponds to the same linear-response $R_V = V_V/I$ as in the maps of Fig. 2 of the main text and *fig. S9*. This is the Knudsen flow regime. At higher currents ($I > 50$ μA), the voltage response becomes nonlinear reaching first a maximum and then changing the sign to negative. This is because the current heats up the electron system and drives it into the Navier-Stokes regime such that whirlpools appear near the injection point. At even higher currents, $V_V$ changes its sign again, from negative to conventional positive, indicating that the electron temperature becomes high enough ($> 300$ K) and the system approaches the high-*T* regime of small $D_v$. If we increased the cryostat temperature to $100$ K (*fig. S8A*), the electron system entered the viscous-flow regime even at vanishingly small probing currents, and the *I-V* curves – linear over a large range of $I$ – yield negative $R_V$, in agreement with the results presented in the main text. At sufficiently high currents, the system again exhibits positive $V_V$, which corresponds to dominating stray currents. Further increase in $T$ in *fig. S10A*, changes the character of *I-V* characteristics once again because the system is already close to the transport regime of small $D_v$, even without being heated by current. Note that these changes are closely connected with the Gurzhi effect reported in *fig. S8*.

However, because the vicinity geometry is much more sensitive to a viscous flow contribution, voltage rather than its derivative changes the sign as a function of $I$ in *fig. S10*.

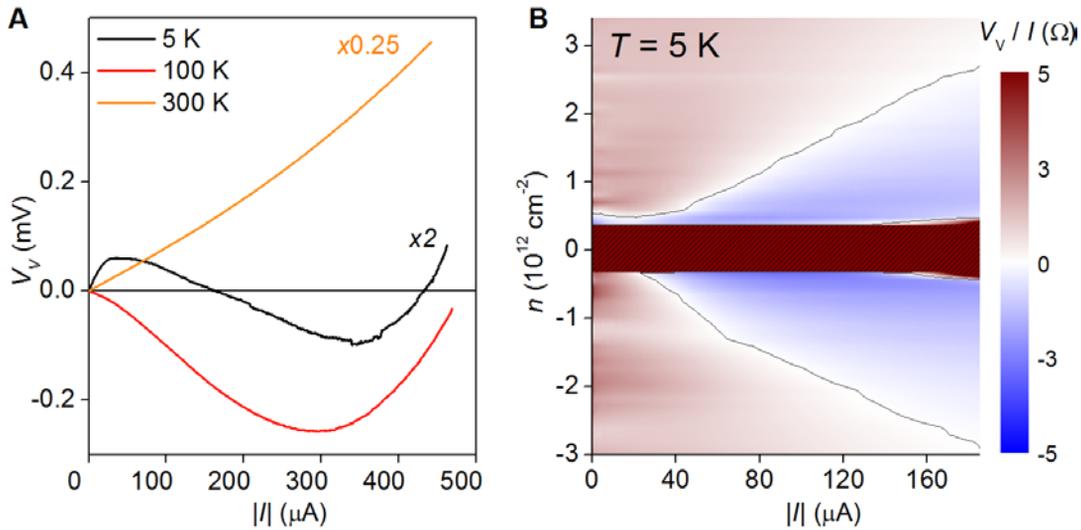

**Fig. S10. Vicinity voltage as a function of applied current.** (**A**) $I$-$V$ characteristics at three characteristic $T$ for a BLG device at hole doping $n = -2.5 \times 10^{12}$ cm$^{-2}$. (**B**) Map of the normalized nonlinear vicinity resistance $V_V/I$ measured at $T = 5$ K. It is important to note that large $I$ can lead to temperature gradients between different contacts and, as a result, spurious thermoelectric signals may appear in such measurements. Because the thermopower contribution depends only on the absolute value of $I$ and not on its sign, the contribution can easily be eliminated by symmetrizing $V_V$ with respect to the direction of dc current. This procedure was applied for the shown plots and, accordingly, they are presented as a function of $|I|$. The brown rectangle in (B) is the region around the CNP with no collected data to avoid overheating and switching between different mesoscopic states.

For further comparison between effects of electron heating and cryostat's $T$, *fig. S10B* shows a map of $V_V/I$, the nonlinear vicinity response normalized by the applied current. This map closely resembles the $R_V(T,n)$ maps in Fig. 2 of the main text and *fig. S9* and also shows a clear transition from normal electron flow at low *T* to backflow at intermediate electron temperatures. Note that in *fig. S10B* we had to limit our measurements to relatively small $I < 200$ μA so that the transport regime dominated by stray currents (approached above 400 μA in *fig. S10A*) could not be reached. This is because such high currents occasionally switched our devices between different mesoscopic states whereas the $V_V$ maps required a few days of continuous measurements. For the same reason, we avoided measurements of $V_V(I)$ around the charge neutrality point in *fig. S10B* where high resistivity of graphene resulted in significant Joule heating even for relatively small currents.

The observed strong enhancement of the negative vicinity signal with increasing the electron temperature is in good agreement with the expected behavior of local whirlpools inside graphene's electron liquid and, also, rules out a contribution from single-particle ballistic effects. Indeed, we found experimentally that the latter phenomena such as negative transfer resistance and magnetic focusing (*19,21*) are rapidly and monotonically suppressed with increasing $I$.

#12 Dependence of electron backflow on distance to the injection contact

We have investigated how negative vicinity resistance decays with increasing the distance $L$ between the injection and voltage contacts (see the sketch in *fig. S11*). This figure shows examples of the temperature dependence of $R_V$ in the linear $I$-$V$ regime for different $L = 1$, $1.3$ and $2.3\,\mu m$, which were measured for the same BLG device at a fixed carrier concentration of $1.5 \times 10^{12}$ cm$^{-2}$. All the plotted curves exhibit negative $R_V$ but the temperature range in which the backflow occurs rapidly narrows with increasing $L$, and we could not detect any backflow for $L > 2.5\,\mu m$ in any of our devices. The magnitude of the negative signal is found to decay rapidly (practically exponentially) with $L$ (top inset of *fig. S11*), yielding a characteristic scale of $\approx 0.5\,\mu m$. This provides a qualitative estimate for the size of electron whirlpools, in agreement with our theoretical estimates for $D_\nu$. Indeed, for the particular device in *fig. S11*, we can estimate $D_\nu \approx 0.4\,\mu m$ using our independent measurements of $\nu \approx 0.1$ m$^2$s$^{-1}$ and $\tau \approx 1.5$ ps (see the main text and above).

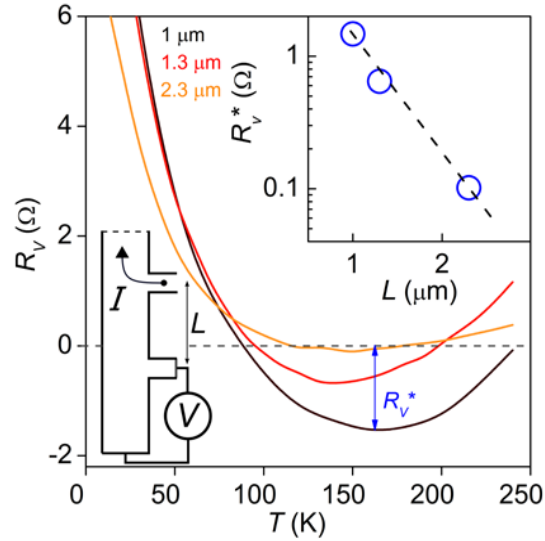

**Fig. S11. Vicinity resistance measured at different distances from the injection contact.** All the contacts were $\approx 0.3\,\mu m$ in width. Top inset: Maximum value of negative $R_V$ as a function of $L$. The dashed curve is the best fit with $D_\nu \approx 0.5\,\mu m$. The probing current was $0.3\,\mu A$.

#13 Stray-current contribution to the vicinity resistance

In the vicinity geometry, stray currents near the voltage probe are not negligible. Their contribution to the measured vicinity resistance is given by the first term $b\sigma_0^{-1}$ in Eq. (4) of the main text where $b$ is the geometrical factor dependent on *L, W* and width of the contact regions (*36*). Fig. 1E of the main text and *fig. S1D* clearly show that the classical contribution was rather significant and competed with the viscous term in $R_V$ over a range of $T$ and $n$. In this report, we have deliberately focused on the sign change in $R_V$ because the negative resistance is an exceptional qualitative effect, which in our case cannot be explained without taking into account a finite viscosity of the electron liquid. However, to elucidate the hydrodynamic behavior in more detail, we can go a step further and analyze the anomalous part of $R_V$, which comes on top of the contribution from stray currents. To this end, we write $\Delta R_V \equiv R_V - b/\sigma_0$ to isolate the second part of Eq. (4) which depends on $D_\nu$ and arises exclusively due to a finite viscosity. *Fig. S12* shows a typical example of $\Delta R_V$ observed in our devices. It is clear that at *T* > 50 K the negative $\Delta R_V$ extends over the entire range of carrier concentrations away from the CNP (*fig. S12A*). *Figure S12B* suggests that electron whirlpools persist well above room *T*.

It is important to note that the above subtraction analysis is based on the assumption of spatially uniform $\sigma_0$ whereas the experimental devices exhibit a certain level of charge inhomogeneity, especially close to the CNP. Qualitatively, one can gauge the influence of charge inhomogeneity from the pronounced electron-hole asymmetry in the $R_V$ maps, which in theory should be symmetric. The asymmetry was found to be contact dependent and arises due to non-uniform charge distribution near the vicinity contacts. Furthermore, the subtraction analysis is not applicable in the low-*T* regime because it ignores single-particle ballistic effects that modify stray currents on a distance of the order of the mean free path. Notwithstanding these limitations, the subtraction procedure in *fig. S12* provides a qualitatively accurate picture, especially at high *T* where single-particle phenomena can be neglected and for $n \gtrsim 1 \times 10^{12}$ cm$^{-2}$ where the electron system become more uniform.

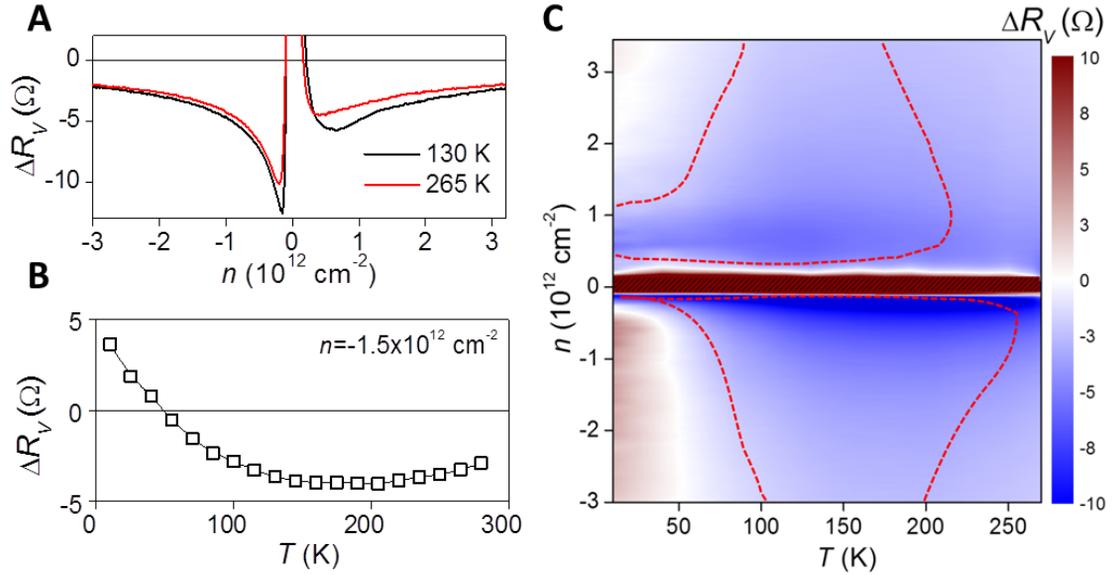

**Fig. S12. Hydrodynamic part of vicinity resistance after subtracting a calculated contribution from stray currents.** (**A**) $\Delta R_V(n)$ at two characteristic temperatures and (**B**) $\Delta R_V(T)$ for a typical carrier density away from the neutrality point. (**C**) Map $\Delta R_V(n,T)$ covering the entire range of measured temperatures and concentrations. Data are for the same device as in Fig. 2B of the main text. The red traces outline the region of negative $R_V$ in Fig. 2B. The brown rectangle indicates the region with $\Delta R_V$ > +10 Ohm around the CNP where our hydrodynamic analysis is not expected to be applicable.

**#14 Ballistic contribution due to reflection from device boundaries**

Charge carriers reflected from the boundary opposite to the current-injecting contact can reach the vicinity probe if $\ell$ is comparable with the travel distance of $\approx 2W$. In this case, one can speculate that reflected electrons can give rise to a contribution similar to the negative bend resistance usually observed for ballistic Hall crosses (*19*). To this end, we have performed numerical simulations using the Landauer-Büttiker formalism and diffusive scattering at graphene edges. The analysis is standard and, therefore, not reported here for the sake of brevity. The simulations yielded the negative bend resistance for the Hall bar geometry, as expected, but we could find only positive contributions for the vicinity geometry. Therefore, the standard theory of ballistic transport cannot explain negative $R_V$. More importantly, our experimental observations also disagree with the above scenario involving ballistic reflection from device boundaries. First, negative $R_V$ is observed for $W$ up to 4 μm and typical $\ell < 2$ μm $< 2W$ so that the number of electrons coming back to the boundary of origin is exponentially small, $\exp(-2W/\ell)$. Second, we have not observed any dependence of the amplitude of negative $R_V$ on $W$, beyond usual variations for different contacts and devices. All our devices showed similar behavior, independent of their size and features such as contacts present at the opposite edge. Third, $\ell$ increases with decreasing $T$ and, therefore, any ballistic contribution is expected to be most pronounced at low $T$. In contrast, $R_V$ is always found positive in the regime of longest $\ell$ (that is, at low $T$ and high $n$), in agreement with our numerical analysis.

Despite the overwhelming evidence described above, let us present an additional set of experiments that further prove little contribution from reflected electrons into $R_V$ and confirm its positive sign. We fabricated devices similar to those described above but submicron slits were added between injecting and vicinity contacts (see *fig. S13*). The basic idea is that such obstacles should stop viscous backflow from reaching the vicinity probe (effective distance $L$ increases significantly). *Fig. S13A* shows that, if no slit is present between the contacts, we observed the standard behavior for $R_V$. It is positive at long $\ell$ at low $T$ but changes sign at higher $T$ becoming most negative around $150 - 200$ K, in agreement with measurements for the other devices (cf. Fig. 2 of the main text and *fig. S9*). On the other hand, if a slit is added next to a vicinity probe, $R_V$ does not change its sign remaining positive (*fig. S13B*). At low $T$, this positive signal is attributed to reflected ballistic electrons reaching the voltage probes for $\ell > W$. Note that $R_V$ in *fig. S13B* is notably smaller than that in *fig. S13A*, in agreement with our numerical simulations and general expectations due to the shadow provided by the slit. As $T$ increases, $R_V$ decreases to zero because for $\ell \ll W$ the geometry gradually becomes nonlocal (*24*).

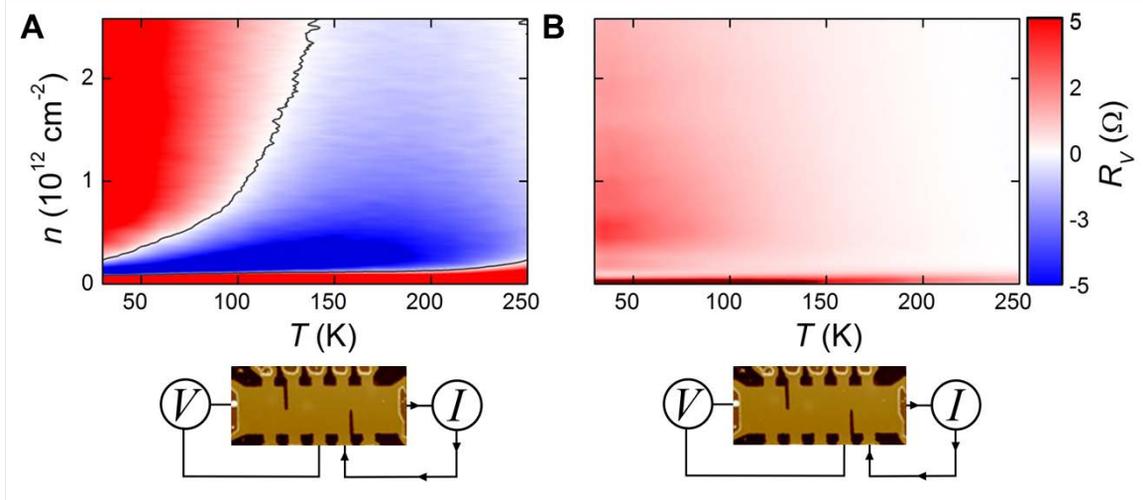

**Fig. S13. Placing an obstacle to shade the vicinity probe.** (**A**) $R_V(n,T)$ for the 'standard'' vicinity geometry as shown schematically in the panel below. The panel utilizes an atomic-force micrograph of a device with two narrow notches etched through encapsulated SLG; $W = 2$ μm. (**B**) Similar measurements but in the geometry where the slit separates vicinity and injector contacts. The schematic is shown in the panel below.

#15 Numerical simulations of hydrodynamic equations

To solve the linearized steady-state Navier-Stokes equation we discretized the differential operators on a square mesh and solve the corresponding sparse linear system. To improve the discretization of the differential operators, the values of the potential and the two components of the velocity are sampled on three different staggered meshes (*37*). The meshes were chosen in such a way that, at the boundary, the velocity component orthogonal to the boundary was sampled.

Different BCs were used to represent the sample edge, an open contact and a current-carrying contact:
(i) the velocity component orthogonal to the edges vanished;
(ii) the total current flowing through an open contact vanished;
(iii) the total current flowing through a current contact was fixed by the experimental conditions.

At the sample boundary (both sample edge and contacts) either no-slip or free-surface boundary conditions were implemented to fix the velocity component parallel to the boundary. We reiterate that the velocity component parallel to the boundary was sampled at a finite distance from the boundary due to the choice of the staggered meshes. An auxiliary set of velocity variables, parallel to the boundary, was introduced just outside of the sample area to implement the desired BCs. More specifically, no-slip boundary conditions were implemented by requiring that the velocity components parallel to the boundary, just inside and outside the sample, were opposite. The free-surface BCs were implemented by discretizing the differential relation of Eq. (*S7*) and making use of the auxiliary set of velocity variables. We used a few hundred mesh nodes in each direction for a typical sample. The solution of the sparse linear system typically takes a few seconds on a desktop computer and its output is the full potential and velocity profile with the desired BC.

Finally, we comment on the dependence of Eq. (4) of the main text on the BCs. We remind the reader that Eq. (4), coupled with a longitudinal four-probe transport measurement of the phenomenological parameter $\tau$, is needed to extract the kinematic viscosity $\nu$ of the electron liquid in graphene. Below we will refer to Eq. (4) of the main text as the 'calibration' curve because it relates the vicinity resistance to the two fundamental parameters of the theory, $\nu$ and $\tau$. Eq. (4) in the main text was derived analytically by assuming free-surface BCs and neglecting a) finite-size effects in the longitudinal Hall bar direction and b) BCs at metal contacts. *Fig. S14* illustrates the dependence of the 'calibration' curve on different BCs, calculated numerically by relaxing constraints a) and b): no-slip (red squares) versus free-surface (filled circles). We also show, for the sake of generality, the analytical result for free-surface BCs, i.e. Eq. (4) of the main text.

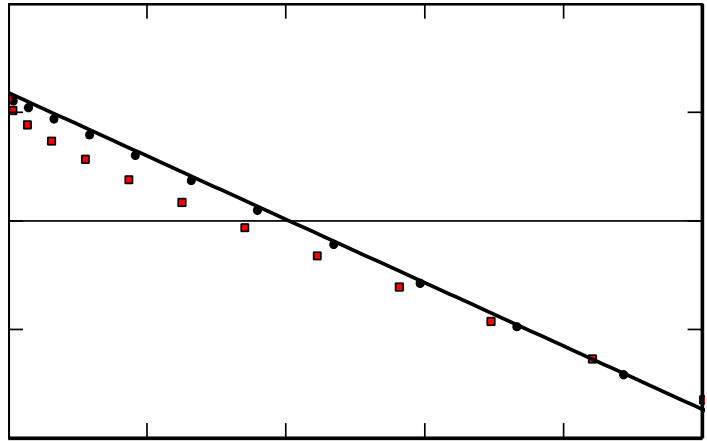

**Fig. S14. 'Calibration' curve for our all-electrical viscometer.** Calculated vicinity resistance $R_V$ (in units of $1/\sigma_0$) as a function of $D_\nu^2$ (in μm²). For $D_\nu^2 \approx 0.2$ μm² the vicinity resistance becomes positive. Filled circles denote fully numerical results for the finite-size BLG device with ten electrodes (obtained by utilizing free-surface BCs). The solid line represents the approximate analytical result of Eq. (4) of the main text, which was obtained with free-surface BCs whereas red squares are the results of numerical calculations obtained by utilizing no-slip BCs ($l_b = 0$).